\documentclass[conference, 10pt]{IEEEtran}

\usepackage{cite}
\usepackage[cmex10]{amsmath}
\usepackage{url}
\usepackage{graphicx}
\usepackage{color}
\usepackage{placeins}
\usepackage{float}
\usepackage{tabularx,colortbl}
\usepackage{ifthen}

\makeatletter

\newcounter{author}
\renewcommand{\author}[2][]{
   \stepcounter{author}
   \@namedef{author@\theauthor}{#2}
   \@namedef{authorlabel@\theauthor}{#1}
}

\newcounter{address}
\newcommand{\address}[2][]{
   \stepcounter{address}
   \@namedef{address@\theaddress}{#2}
   \@namedef{addresslabel@\theaddress}{#1}
}

\newcommand{\alsep}{and}

\def\newmaketitle{\par%
  \begingroup%
  \normalfont%
  \def\thefootnote{}
  \def\footnotemark{}
  \let\@makefnmark\relax
  \footnotesize
  \footnotesep 0.7\baselineskip
  \normalsize%
  \twocolumn[\thenewmaketitle\@IEEEaftertitletext]%
  \if@IEEEusingpubid
     \enlargethispage{-\@IEEEpubidpullup}%
  \fi
  \endgroup
  \setcounter{footnote}{0}\let\maketitle\relax\let\@maketitle\relax
  \gdef\@thanks{}%
  \let\thanks\relax}

\def\thenewmaketitle{
  \newpage
  \begin{center}%
    \vskip0.2em{\Huge\@IEEEcompsoconly{\sffamily}\@IEEEcompsocconfonly{\normalfont\normalsize\vskip 2\@IEEEnormalsizeunitybaselineskip
   \bfseries\large}\@title\par}\vskip1.0em\par%
    \vspace{1ex}
    \newcounter{c@author}
    \newcounter{c@tmp}
    \ifthenelse{\value{author}=2}{%
      \newcommand{\liand}{ and }}{%
      \newcommand{\liand}{, and }}
    \ifthenelse{\value{address}<2}{%
      \@nameuse{author@1}%
      \stepcounter{c@author}%
      \whiledo{\value{c@author}<\value{author}}{%
        \setcounter{c@tmp}{\value{author}}%
        \addtocounter{c@tmp}{-\value{c@author}}%
        \ifthenelse{\value{c@tmp}=1}{%
          \renewcommand{\alsep}{\liand}}{\renewcommand{\alsep}{, }}%
        \stepcounter{c@author}\alsep \@nameuse{author@\thec@author}}\\%
    }
    {
      \@nameuse{author@1}${}^{(\ref{\@nameuse{authorlabel@1}})}$%
      \stepcounter{c@author}%
      \whiledo{\value{c@author}<\value{author}}{%
      \setcounter{c@tmp}{\value{author}}%
      \addtocounter{c@tmp}{-\value{c@author}}%
      \ifthenelse{\value{c@tmp}=1}{%
        \renewcommand{\alsep}{\liand}}{\renewcommand{\alsep}{, }}%
      \stepcounter{c@author}\alsep \@nameuse{author@\thec@author}%
        ${}^{(\ref{\@nameuse{authorlabel@\thec@author}})}$%
      }
    }
    \vspace{0.2ex}

    \ifthenelse{\value{address}>0}{%
      \ifthenelse{\value{address}=1}{
        {\@nameuse{address@1}}
      }
      {
        \newcounter{c@address}

        \begin{center}
        \whiledo{\value{c@address}<\value{address}}
        {
          \refstepcounter{c@address}
            ${}^{(\thec@address)}$\,%
              \label{\@nameuse{addresslabel@\thec@address}}%
              \@nameuse{address@\thec@address}\\ %
        }
        \end{center}
      } 
    }
    {
      \relax
    }
  \end{center}
}

\makeatother

\title{Engineering Spatial Dispersion to Synthesize Arbitrary Spatial Filters Based on Metagratings}

\author[org1]{Jinyong Kim}
\author[org2]{Minseok Kim\textsuperscript{*}}

\address[org1]{School of Electronic and Electrical Engineering, Hongik University, 94 Wausan-ro, Mapo-gu, Seoul, 121-791, Korea \\ \textsuperscript{*}minseok.kim@hongik.ac.kr}

\begin{document}

\newmaketitle

\begin{abstract}
This paper presents a design framework for synthesizing angularly selective spatial filters using non-uniform metagratings. While traditional metagratings focus on channeling energy into higher-order Floquet modes for a fixed incidence angle, we leverage the fundamental mode as a versatile degree of freedom to engineer spatial dispersion over a continuous angular spectrum. By strategically distributing non-uniformly loaded metallic wires and rigorously modeling their mutual interactions through an impedance-matrix formulation, we realize prescribed angular transfer functions with high efficiency. In particular, the framework is validated at 3.5 GHz through full-wave simulations of (i) low-pass, (ii) high-pass, and (iii) all-pass spatial filters.  The results demonstrate that fundamental-mode engineering in non-uniform metagratins offers a highly efficient platform for advanced spatial wave manipulation.

\end{abstract}


\section{Introduction}

Metagratings have emerged as a powerful platform for wavefront engineering by employing sparse arrangements of subwavelength unit cells that precisely control diffraction modes via the Floquet-Bloch theory. This framework is particularly robust because it natively accounts for mutual coupling—and thus the resulting non-local responses—between unit cells, enabling wave-manipulation efficiencies that often surpass those of dense metasurfaces. Despite this inherent ability to model non-local interactions, most metagratings are designed for a fixed incidence angle to channel incident electromagnetic (EM) power into specific higher-order Floquet modes, performing tasks such as anomalous reflection, refraction, or beam splitting~\cite{Ra'di2017PRL, Popov2018PRA}. Consequently, the broader potential for a prescribed, angle-selective response remains largely unexplored that would allow the surface to exhibit distinct EM scattering responses depending on the direction of the impinging incident wave.

To address this missing functionality, recent studies have proposed engineering the spatial dispersion of metasurfaces. Unlike metagratings, these surfaces comprise an array of unit cells with deep subwavelength periodicity, allowing them to be modeled as homogenized, zero-thickness sheets. These sheets are typically characterized by effective surface parameters such as polarizability, susceptibility, and impedance, which are rigorously related to the incident and scattered fields through generalized sheet transition conditions (GSTCs). Notable analytical frameworks, including multipolar~\cite{Achouri2022IEEEAPS} and Padé expansions~\cite{Rahmeier2023IEEETAP}, have been introduced to extend GSTCs to properly model spatial dispersion and accurately characterize the scattering response as a function of incidence angles. Nevertheless, these methods remain primarily within the theoretical domain, as a general physical realization for unit-cell synthesis has yet to be reported.

Following these theoretical advancements, a practical implementation strategy based on multi-layered metasurfaces has been proposed~\cite{Shaham2025IEEETAP}. This approach directly maps the tangential and normal surface susceptibilities required for a targeted angular selectivity to the specific surface impedances that comprise the multi-layered architecture, thereby providing a physical route to induce spatial dispersion~\cite{Shaham2025IEEETAP}. Nevertheless, these multi-layered metasurfaces remain largely restricted to uniform structures composed of identical unit cells. This limitation stems from the significant analytical complexity required to model non-local effects in non-uniform (inhomogeneous) metasurfaces, where the spatial variation of susceptibilities must be rigorously accounted for~\cite{Dugan2023IEEETAP}. As a result, only highly specific and relatively simple angular responses, such as all-pass characteristics, have been demonstrated to date~\cite{Mustafa2025PRB}.

In light of this, it is evident that there is a clear gap in the realization of a surface—whether in the form of a metagrating or metasurface—that would allow synthesizing arbitrary scattering responses as a function of the incidence angle. In this work, we address this gap by utilizing a non-uniform metagrating and leveraging its fundamental mode as a degree of freedom for engineering spatial dispersion. This stands in contrast to conventional metagrating analyses, which primarily focus on channeling incident power into higher-order Floquet modes at a fixed incidence angle. Utilizing this approach, we report various metagrating-based spatial filters that offer angular-selective responses beyond the capabilities of uniform spatially-dispersive metasurfaces. The versatility of our approach is validated through full-wave simulations of three design examples operating at 3.5 GHz: (i) low-pass, (ii) high-pass, and (iii) all-pass spatial filters whose simulation results show excellent agreement with the analytical prediction.

\section{CONCEPT AND DESIGN FRAMEWORK}
\subsection{Geometry of proposed supercell}
\begin{figure}[t!]
\begin{center}
\noindent
  \includegraphics[width=0.9\columnwidth]{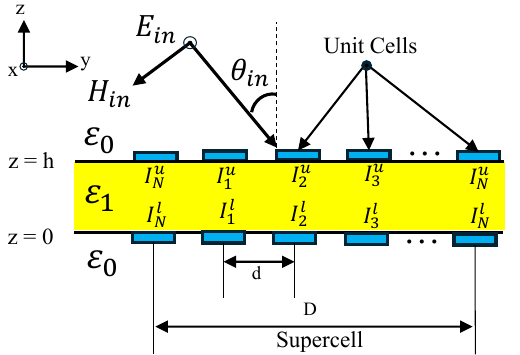}
  \caption{Front view of proposed spatial filter}
  \label{fig:OverallSchematic}
\end{center}
\end{figure}

Fig.~\ref{fig:OverallSchematic} illustrates the overall geometry of the proposed two-layer metagrating that functions as a spatial filter.
The metagrating operates at 3.5 GHz ($\lambda_0 \approx 85.7$ mm) and consists of two parallel arrays of unit cells that are realized as capacitively-loaded metallic wires with a width of $w = 0.762$ mm ($\approx 0.009\lambda_0$). It is envisioned that the arrays are formed by periodic repetition of supercells comprising $N$ wires with different capacitive loading and a uniform spacing of $d = D/7$. The number of unit cells within each supercell, $N$, is determined by the complexity of the target angular response. Specifically, $N=4$ is utilized for the low-pass and all-angle transmission responses. On the other hand, for the high-pass response, $N$ is increased to 5 to enhance the design flexibility. The supercells are separated by a $h = 7.874$ mm-thick ($\approx 0.092\lambda_0$) Rogers RT/Duroid 5880 substrate ($\epsilon_r = 2.2$, $\tan \delta = 0.0009$), with the entire structure embedded in free space ($\epsilon_0$). A transverse-electric (TE) plane wave is assumed, which impinges the surface from the upper region ($z > h$) at an angle $\theta_{in}$, varying from $0^\circ$ to $60^\circ$.

To suitably select the periodicity of our supercell, $D$, we invoke the Floquet-Bloch theorem. According to the theorem, the modes that propagate in free space for a certain period $D$ and incidence angle $\theta_{in}$ satisfy the following condition:
\begin{equation}
-\left\lfloor \frac{D}{\lambda_0}(1 + \sin \theta_{in}) \right\rfloor \le m \le \left\lfloor \frac{D}{\lambda_0}(1 - \sin \theta_{in}) \right\rfloor \quad 
\end{equation}    
where the symbol $\lfloor \cdot \rfloor$ denotes the floor function. Based on this relation, we select a supercell period of $D = 0.9\lambda_0$ such that the structure supports a maximum of only two propagating modes within the range of $\{0^\circ, 60^\circ\}$. While a larger periodicity could be adopted to create a sparser metagrating for easier fabrication, allowing additional propagating modes would lead to unwanted power leakage into higher-order diffraction channels. Such leakage inevitably degrades both the filtering efficiency and the precision of the fundamental mode ($m=0$) control, which is the primary objective of this study.

\subsection{Numerical Formulation of the Proposed Metagrating}
Based on the established geometric configuration, we present a systematic design framework to synthesize (i) low-pass, (ii) high-pass, and (iii) all-pass spatial filters. While the design framework is consistent with the one reported in Ref.~\cite{Casolaro2020TAP}, we summarize its key steps to provide a context for the optimization strategy developed herein. It is important to emphasize that, although the present work and Ref.~\cite{Casolaro2020TAP} both leverage the Floquet-Bloch theorem, their physical objectives and operational principles are fundamentally distinct. Specifically, the approach in Ref.~\cite{Casolaro2020TAP} focuses on anomalous reflection, refraction, and beam splitting at a single, fixed incidence angle, where the objective is to redirect incident energy into specific higher-order Floquet modes. In contrast, the present work aims to engineer the angular transfer function over a continuous spectrum of incidence angles (from $0^\circ$ to $60^\circ$). Instead of scattering energy into higher-order modes, our design selectively redirects incident waves into the fundamental ($0$th-order) mode as either a reflected or transmitted wave, depending on the angle of incidence. By determining the specific incidence angles at which the metagrating reflects or transmits, we realize the desired spatial filters.

To this end, we aim to accurately evaluate the induced current distributions on each wire required to achieve the desired spatial filtering responses. We first model the entire metagrating structure as a $2N \times 2N$ impedance system, $\mathbf{Z}_\text{MG}$, that directly relates the external incident field and the induced currents as,
\begin{equation}
\mathbf{Z}_\text{MG}\mathbf{I}=(\mathbf{Z}_\text{self} + \mathbf{Z}_\text{mutual} + \mathbf{Z}_\text{load}) \mathbf{I} = \mathbf{V}
\label{eq:ZI_V}
\end{equation}
\noindent where $\mathbf{Z}_\text{MG}$ is decomposed into three components: (i) $\mathbf{Z}_\text{self}$, which models the self-coupling of each wire with its own radiation, (ii) $\mathbf{Z}_\text{mutual}$, which accounts for the mutual coupling between all unit cells within the arrays, and (iii) $\mathbf{Z}_\text{load}$, a diagonal matrix containing the specific wire impedances (capacitive loading) for each unit cell. In Eq.~\eqref{eq:ZI_V}, $\mathbf{I}$ and $\mathbf{V}$ are $2N \times 1$ vectors representing the unknown induced currents and the external excitation at the wire positions, respectively. The excitation vector $\mathbf{V}$ can be readily obtained from the incident electric field values evaluated at each wire location $(y_n, z_n)$ in the absence of the metallic wires, defined as $V_n = E_{inc}(y_n, z_n)$.

Provided that $\mathbf{Z}_\text{MG}$ is fully defined, Eq.~\eqref{eq:ZI_V} ensures that the unknown induced currents can be analytically determined. Since $\mathbf{Z}_\text{self}$ and $\mathbf{Z}_\text{mutual}$ account for the self- and mutual-coupling effects dictated solely by the metagrating geometry, they remain constant for a fixed configuration. Consequently, the task of synthesizing a specific induced current distribution is effectively reduced to the determination of the individual load impedances for each wire (i.e., $\mathbf{Z}_\text{load}$). Within this framework, the reactive elements in $\mathbf{Z}_\text{load}$ serve as the design variables, which are optimized to satisfy the target angular response. The self-coupling matrix is defined as $\mathbf{Z}_\text{self} = Z_{s} \mathbf{U}$, where $\mathbf{U}$ is a $2N \times 2N$ identity matrix and $Z_{s} = \frac{\eta_0 k_0}{4} H_0^{(2)}(k_0 r_\text{eff})$ represents the self-impedance of an individual wire with an effective radius $r_\text{eff} = w/4$. In addtion, the mutual coupling impedance between the wires is given by, 
\begin{equation}
\mathbf{Z}_\text{mutual} = \begin{pmatrix} \mathbf{Z}^\text{on} & \mathbf{Z}^\text{off} \\ \mathbf{Z}^\text{off} & \mathbf{Z}^\text{on} \end{pmatrix}
\end{equation}
\noindent where the sub-matrices $\mathbf{Z}^\text{on}$ and $\mathbf{Z}^\text{off}$ account for the mutual interactions between unit cells located on the same plane and across different planes, respectively. These sub-matrices are given by~\cite{Casolaro2020TAP},
\begin{subequations}
\begin{align}
Z_{qp}^{\text{off}} &= \frac{\eta_0 k_0}{2D} \sum_{m=-\infty}^{\infty} \frac{T_m}{\beta_{z,m,0}} e^{jk_{ym}(p-q)d} \\
Z_{qp}^{\text{on}} &= \begin{cases} 
\displaystyle \frac{\eta_0 k_0}{2D} \sum_{m=-\infty}^{\infty} \frac{(1 + R_m)}{\beta_{z,m,0}} e^{jk_{ym}(p-q)d}, & p \neq q \\[15pt]
\begin{aligned}
&\frac{\eta_0 k_0 (1 + R_0)}{2D \beta_{z,0,0}} - \frac{\eta_0 k_0}{4} \\
&+ j \frac{\eta_0 k_0}{2\pi} \left[ \log \left( \frac{k_0 D}{4\pi} \right) + \gamma \right] \\
&+ \frac{\eta_0 k_0}{2D} \sum_{\substack{m=-\infty \\ m \neq 0}}^{\infty} \left[ \frac{1 + R_m}{\beta_{z,m,0}} - \frac{j D}{2\pi |m|} \right],
\end{aligned} & p = q
\end{cases}
\end{align}
\end{subequations}
\noindent where $\gamma \approx 0.577$ is Euler-Mascheroni constant, $\eta_0$ is the free-space impedance, and $k_0$ is the free-space wavenumber. The indices $p$ and $q$ denote the ordinal positions of the wires within the supercell, representing the source and observation points, respectively. The transverse wavenumber of the $m$-th diffraction mode is defined as $k_{y,m} = k_0 \sin\theta_{in} + 2\pi m/D$, while $\beta_{z,m,0} = \sqrt{k_0^2 - k_{y,m}^2}$ and $\beta_{z,m,1} = \sqrt{k_1^2 - k_{y,m}^2}$ represent the corresponding longitudinal wavenumbers in the free-space (subscript '0') and dielectric (subscript '1') regions. Here, $k_1 = k_0\sqrt{\epsilon_r}$ denotes the wavenumber within the dielectric substrate. Finally, $R_m$ and $T_m$ are the reflection and transmission coefficients of the background structure for the $m$-th mode~\cite{Casolaro2020TAP}.

\section{Optimization of Metagratings and Full-Wave Results}
Having established the relationship between $\mathbf{V}$ and $\mathbf{I}$ via the impedance matrices, we employ the particle swarm optimization (PSO) method to determine the optimal configuration of $\mathbf{Z}_\text{load}$ that induces the currents necessary to achieve the desired spatial filtering response. Specifically, for a given trial set of $\mathbf{Z}_\text{load}$, the induced currents are analytically calculated via Eq.~\eqref{eq:ZI_V}, from which the power coupled to the $0$th-order (fundamental) mode is evaluated over a range of incidence angles from $0^\circ$ to $60^\circ$. Letting $A_0$ and $B_0$ denote the reflection and transmission amplitudes of the $0$th-order mode, respectively, they are evaluated as~\cite{Casolaro2020TAP},
\begin{subequations}
    \begin{align}
A_0 &= -\frac{k_0 e^{j \beta_{z,0,0} h}}{\beta_{z,0,0}} \left[ (1 + R_0) \Psi_0^u + T_0 \Psi_0^l \right] +  E_0 R_0 e^{2j \beta_{z,0,0} h} \\
B_0 &= -\frac{k_0}{\beta_{z,0,0}} \left[ T_0 \Psi_0^u + (1 + R_0) \Psi_0^l \right] +  E_0 T_0 e^{j \beta_{z,0,0} h}  
    \end{align}
    \label{eq:A0_B0}
\end{subequations}
\noindent where the supercript $u$ and $l$ denote upper ($z=h$) and lower ($z=0$) arrays, respectively. $\Psi_0^{u(l)}$ is related to the induced currents as~\cite{Casolaro2020TAP},
\begin{equation}
\Psi_0^{u(l)} = \sum_{q=1}^N \frac{\eta_0 I_q^{u(l)}}{2D} e^{j k_{y0} (q-1)d}.    
\end{equation}
\noindent The PSO algorithm then aims to minimize the cost function, $C$, defined as:
\begin{equation}
C = \sum_{\theta = 0^\circ}^{60^\circ} \left[ \left( \frac{|A_0|^2}{|E_0|^2} - \xi_\text{target}^\text{ref}\right)^2 + \left(\frac{|B_0|^2}{|E_0|^2} - \xi_\text{target}^\text{tr}\right)^2 \right]
\end{equation}
\noindent where $\xi_\text{target}^\text{ref}$ and $\xi_\text{target}^\text{tr}$ represent the desired angular responses in the reflection and transmission domain, respectively. Following the PSO, a gradient-based refinement is performed using \texttt{fmincon}, where the optimal set identified by the PSO serves as the initial starting point. During both optimization stages, strict physical constraints are imposed to ensure the proposed metagrating remains purely passive and lossless. For this purpose, the real part of the load impedance is set to zero ($Re(Z_\text{load}) = 0$). Additionally, the search space for the capacitive loads is constrained to the range or $-j600 \, \Omega \le Im(Z_\text{load}) \le -j50 \, \Omega$. Table \ref{tab:optimized_impedance} summarizes the optimized capacitive load impedances for low-pass, high-pass, and all-pass spatial filters.
\begin{table}[t!]
    \centering
    \caption{OPTIMIZED IMPEDANCE OF SPATIAL FILTER($\Omega$)} 
    \label{tab:optimized_impedance}
    \renewcommand{\arraystretch}{1.3}
    \begin{tabular}{c || c c c}
    \hline
        $Z_{\text{load}}$ & Low Pass & High Pass & All-angle \\ \hline
        $Z_1$    & $-j300$    & $-j551.4$  & $-j339.52$ \\
        $Z_2$    & $-j166.7$  & $-j162.65$ & $-j116.8$  \\
        $Z_3$    & $-j163.8$  & $-j600$    & $-j500.00$ \\
        $Z_4$    & $-j188.35$ & $-j390.99$ & $-j197.97$ \\
        $Z_5$    & $-j300$    & $-j169.89$ & $-j500.00$ \\
        $Z_6$    & $-j300$    & $-j218.58$ & $-j208.65$ \\
        $Z_7$    & $-j205.03$ & $-j142.6$  & $-j248.95$ \\
        $Z_8$    & $-j300$    & $-j600$    & $-j117.69$ \\
        $Z_9$    &            & $-j226.69$ &            \\
        $Z_{10}$ &            & $-j138.06$ &            \\ \hline
    \end{tabular}
\end{table}

\begin{figure}[t!]
\begin{center}
\noindent
  \includegraphics[width=0.9\columnwidth]{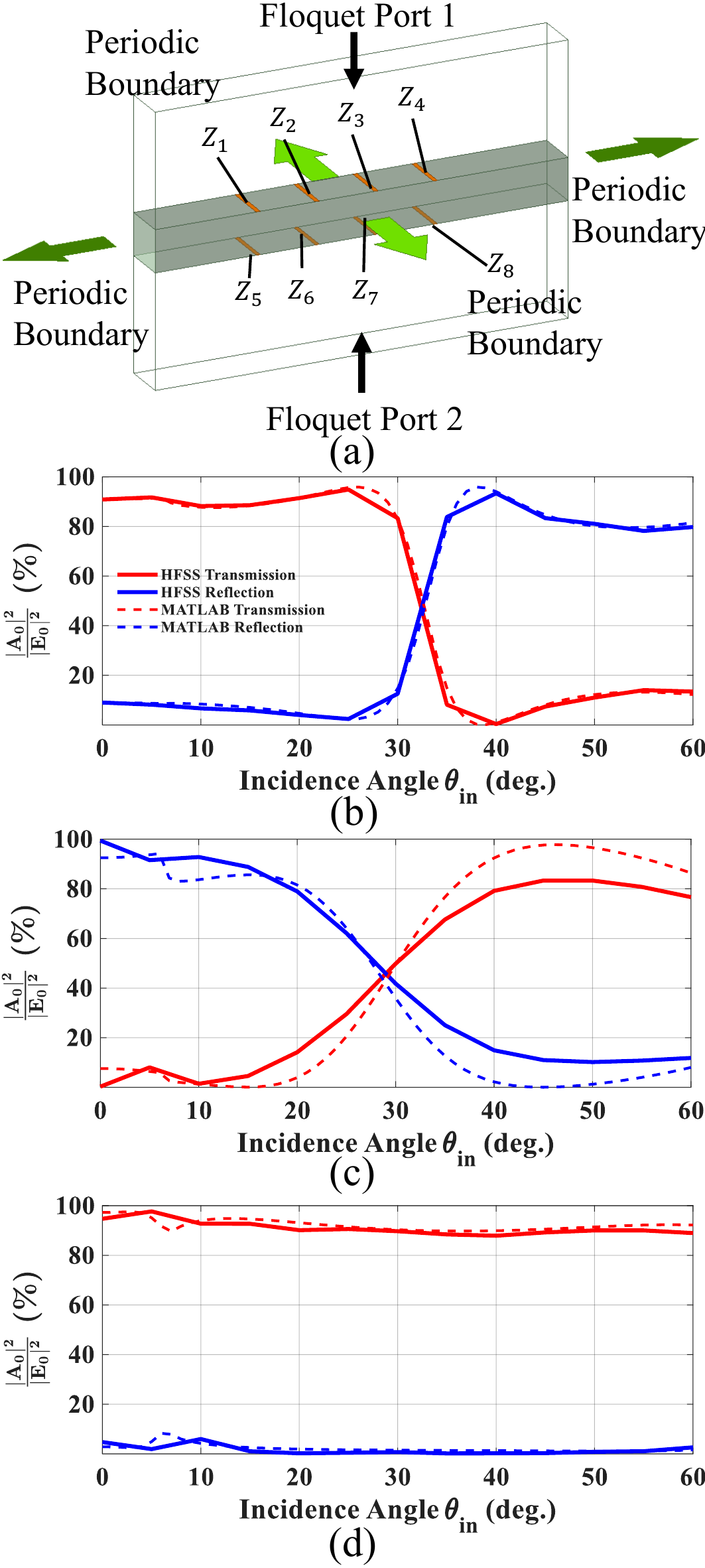}
  \caption{(a) Schematic of simulation setup employing impedance boundaries that implement the optimized results for wires. (b)-(d) Calculated and simulated 0th-order diffraction efficiencies of the proposed metagrating-based spatial filters at 3.5 GHz: (b) Low-pass response  (c) high-pass response, and (d) wide-angle transmission response with a stable efficiency profile}\label{fig:SimVerifications}
\end{center}
\end{figure}

With the optimized load impedances, the corresponding spatial filtering responses are numerically validated via \texttt{ANSYS HFSS}. Fig.~\ref{fig:SimVerifications}(a) shows the simulation setup used to verify low-pass and all-pass filters where there are eight impedance strips that implement the optimized load impedances. It is noted that an analogous simulation setting is employed with two additional impedance strips to accommodate the high-pass filter simulation. Figs.~\ref{fig:SimVerifications}(b-d) show the simulation results from which it is clearly seen that the the angular responses exhibit the desired low-pass, high-pass, and all-pass responses, respectively. It is noted that the dotted lines in the figures represent the analytical results (obtained via Eq.~\eqref{eq:A0_B0}), which show good agreement compared to the \texttt{ANSYS HFSS} results across the incident angles ranging from $0^\circ$ to $60^\circ$, thereby validating the proposed synthesis framework.
Specifically, for the low-pass filter synthesis, incidence angles below $30^\circ$ are set to be transmitted (Fig.~\ref{fig:SimVerifications}(b)). Both analytical predictions and full-wave simulations demonstrate a sharp roll-off immediately beyond this prescribed $30^\circ$ threshold, while maintaining high transmittance at lower angles. Similarly, the high-pass filter was designed with a critical transition at $30^\circ$, where a distinct inversion between reflection and transmission occurs. This prescribed behavior is clearly observed in Fig.~\ref{fig:SimVerifications}(c) for both analytical predictions and simulation results. Finally, Fig.~\ref{fig:SimVerifications}(d) confirms the successful realization of the all-pass spatial filter, showing excellent agreement between the proposed framework and full-wave simulation.

This level of angular selectivity is particularly noteworthy given that it is achieved using only two layers of sparse unit-cell arrays. For uniform non-local metasurface, achieving similar response with the same number of layers would be extremely challenging, if not impossible~\cite{Mustafa2025PRB}. In contrast, by strategically and non-uniformly distributing discrete wires with varying loads within a supercell, the proposed approach provides a significantly enhanced degree of freedom. This allows for the synthesis of sharp filtering characteristics that would otherwise require a much greater number of layers in uniform architectures. 

\section{Conclusion}

We have demonstrated a robust framework for synthesizing advanced spatial filters based on non-uniform metagratings. By shifting the focus to the engineering of the fundamental mode’s spatial dispersion, we have enabled metagratings to respond selectively to incident waves based on their angle of arrival. Our analytical model, which incorporates non-local mutual coupling, allows for the precise determination of the capacitive loads required to achieve user-defined angular responses. Full-wave simulations for low-pass, high-pass, and all-pass filters show excellent agreement with theoretical predictions, confirming the accuracy of the proposed formulation. The ability to achieve sharp angular selectivity with only two layers underscores the structural efficiency of the non-uniform metagrating approach over uniform non-local metasurfaces, paving the way for the development of compact, high-performance spatial filters for next-generation communication and sensing systems.


\section*{ACKNOWLEDGEMENT}
This work was supported by the National Research Foundation of Korea(NRF) grant funded by the Korea government(MSIT)(RS-2024-00341191).

\bibliographystyle{IEEEtran}
\bibliography{MinseokKim_Refs.bib}

\end{document}